\documentclass[12pt]{article}

\usepackage{lscape}
\usepackage{multirow}
\usepackage{amssymb,amsmath}
\usepackage{amstext}
\usepackage{amscd}
\usepackage{graphics}
\usepackage{epsfig}
\usepackage{shadow}

\def\be{\begin{equation}}
\def\ee{\end{equation}}
\def\beq{\begin{equation}}
\def\eeq{\end{equation}}
\def\bea{\begin{eqnarray}}
\def\eea{\end{eqnarray}} 

\def\eqn#1{(\ref{#1})}

\def\sideremark#1{\ifvmode\leavevmode\fi\vadjust{\vbox to0pt{\vss
 \hbox to 0pt{\hskip\hsize\hskip1em
 \vbox{\hsize3cm\tiny\raggedright\pretolerance10000
  \noindent #1\hfill}\hss}\vbox to8pt{\vfil}\vss}}}

\begin{document}
\thispagestyle{empty}

\vspace{.8cm}
\setcounter{footnote}{0}
\begin{center}
\vspace{-25mm}
{\Large
 {\bf $(p,q)$-form K\"ahler Electromagnetism}\\[5mm]

 {\sc \small
     D.~Cherney$^{\mathfrak C}$, E.~Latini$^{\mathfrak L}$,  and A.~Waldron$^{\mathfrak W}$\\[4mm]

            {\em\small${}^{\mathfrak C,\mathfrak W}\!$
            Department of Mathematics\\ 
            University of California,
            Davis CA 95616, USA\\
            {\tt cherney,wally@math.ucdavis.edu}\\[2mm]
           ${}^\mathfrak{L}$ 
Dipartimento di Fisica,\\ Universit\`a
a di Bologna, via Irnerio 46, I-40126 Bologna, Italy 
and INFN, Laboratori Nazionali di Frascati, CP 13, I-00044 Frascati, Italy \\
{\tt latini@lnf.infn.it}\\[2mm]

            }}

 }

\bigskip

{\sc Abstract}\\[-4mm]
\end{center}

{\small
\begin{quote}

We present a gauge invariant generalization of Maxwell's equations and $p$-form electromagnetism to K\"ahler spacetimes.

\end{quote}
}

\noindent
In their simplest form our K\"ahler Maxwell's equations are given by
$$
\omega^{\bar \jmath i} \Big[\partial_{i}\partial_{\bar \jmath} A_{i_{1}\ldots i_{p}\bar \jmath_{1}\ldots\bar \jmath_{q}}
-p\, \partial_{[i_{1}}\partial_{\bar\jmath}A_{|i|i_{2}\ldots i_{p}]\bar\jmath_{1}\ldots\bar \jmath_{q}}
-q\, \partial_{i}\partial_{[\bar\jmath_{1}}A_{i_{1}\ldots i_{p}|\bar\jmath|\bar\jmath_{2}\ldots\bar \jmath_{q}]}\Big.
\hspace{2cm}
$$
\be
\hspace{7cm}
\Big.
+pq\, \partial_{[i_{1}}\partial_{[\bar \jmath_{1}} A_{|i|i_{2}\ldots i_{p}]|\bar\jmath|\bar \jmath_{2}\ldots\bar \jmath_{q}]}
\Big]=0\, .\label{pq}
\ee
Here $\omega_{i\bar\jmath}$ is the K\"ahler form and we have raised and lowered indices with the K\"ahler metric,
while the dynamical field $A_{i_{1}\ldots i_{p}\bar \jmath_{1}\ldots\bar \jmath_{q}}$ is a $(p,q)$-form on a K\"ahler manifold $M$ of any
complex dimension~$n$. Therefore we call this system of equations  $(p,q)$-form K\"ahler electromagnetism. To understand this system and its gauge 
invariances better we first streamline our notations using some basic Hodge-Lefschetz theory (see, for example~\cite{Griff} for a detailed account and original references).

Firstly we denote the Dolbeault decomposition of the exterior derivative $d$ and the codifferential $d^{*}=\star d\star$ in the usual way
\be
d=\partial + \bar \partial\, ,\qquad d^{*}= \delta + \bar \delta\, ,
\ee
subject to the ${\cal N}=4$ supersymmetry algebra
\be
\{\partial,\delta\}=\frac12 \, \Delta= \{\bar\partial,\bar\delta\}\, ,\label{Q}
\ee 
where $\Delta$ is the form Laplacian. The $R$-symmetries of this superalgebra include the Hodge-Lefschetz  $sl(2)$ algebra
generated by $\{\Lambda,H,L\}$. The operator $\Lambda$ is essentially the contraction by the K\"ahler form appearing in~\eqn{pq},~$H$~returns $p+q-n$ on $(p,q)$ forms on an $n$-dimensional complex manifold while~$L$ is the adjoint of~$\Lambda$ 
with respect to the Hodge inner product. All these operators may be realized in first quantization as symmetries of a quantum mechanical 
${\cal N}=4$ supersymmetric sigma model\footnote{Supersymmetric mechanics on  K\"ahler manifolds has been extensively studied in~\cite{Bellucci:2001ax}.} (see~\cite{FigueroaO'Farrill:1997ks} for an elegant account and history of this theory).
We then have the algebra
$$
[H,\Lambda]=-2\Lambda\, ,\qquad [H,L]=2L\, ,\qquad [\Lambda,L]=H\, ,
$$
\be
[\Lambda,\partial]=\bar\delta\, ,\quad [\Lambda,\bar \partial]=-\delta\, ,\qquad [L,\delta]=\bar \partial\, ,\qquad [L,\bar\delta]=-\partial\label{R}
\, .\ee
In these terms our $(p,q)$-form K\"ahler Maxwell's equations become simply
\begin{center}
\shabox{$\Lambda\,  \partial\bar\partial A =0\, ,$}
\end{center}
\vspace{-.5cm}
\be
\ee
for any differential form $A$ on a K\"ahler manifold $M$ (it is no longer necessary to specify a definite degree for $A$).

These equations share some remarkable similarities with both Maxwell's equations and the linearized Einstein equations.
To see this we must briefly recall the theory of symmetric forms. A symmetric form is any symmetric tensor field compactly expressed
as a function of commuting differentials~$dx^{\mu}$ (for example the metric tensor $ds^{2}= dx^{\mu} g_{\mu\nu} dx^{\nu}$).
On symmetric forms one has useful geometric operations such as the symmetrized gradient ${\bf grad}$, divergence ${\bf div}$
and trace ${\bf tr}$ (see~\cite{symm} for details). In these notations, in a flat background, many physical systems
are generated by the equations
\be
\Big[\square - {\bf grad}\ {\bf div} + \frac 12\,  {\bf grad}^{2} \ {\bf tr}\Big] \varphi=0= {\bf tr}^{2}\varphi\, .\label{einst}
\ee 
For example, if the symmetric form~$\varphi$ is a vector  these are Maxwell's equations {\it in vacua}.
When $\varphi=h_{\mu\nu}dx^{\mu} dx^{\nu}$---metric fluctuations---the above equations are the linearized Einstein equations.
Using the superalgebra~(\ref{Q},\ref{R}), our K\"ahler equations mimic these massless higher spin equations
\be
\Big[\Delta - 2 \partial \delta -2 \bar\partial\bar\delta+2\partial\bar\partial\, \Lambda\Big] A = 0\, .\label{maxeom}
\ee
where $(\partial,\bar\partial)$ play the {\it r\^ole} of ${\bf grad}$, $(\delta,\bar\delta)$ of ${\bf div}$ and the K\"ahler form contraction~$\Lambda$ of~${\bf tr}$. For scalars, the equations~\eqn{maxeom} reduce to the Laplace equation.  In that case, in four real dimensions, viewing
$F=\partial \bar\partial\varphi $ as an abelian field strength, the resulting equation $\Lambda F=0$ is the anti-self-duality condition of~\cite{Itoh}. 
For a $(1,0)$-form $A=A_{i} dz^{i}$
we obtain simply $\delta \partial A =0$, the natural complex generalization of the Maxwell system. {\it I.e.}, for vectors we obtain a holomorphic
and anti-holomorphic copy of Maxwell's equations\footnote{Holomorphic topological Yang--Mills theory was studied in~\cite{Galperin:1990nm,Park}.}. Note, our system is not equivalent to the standard equations $d^{*} d  A=0$.

Our K\"ahler equations clearly enjoy gauge invariances
\be
A\sim A + \partial \alpha + \bar \partial \bar \alpha~\label{gauge}
\ee
for arbitrary forms $\alpha$ and $\bar \alpha$. A corresponding pair of Bianchi identities hold thank to the operator facts
\be
[\delta -\frac12 \partial \Lambda]\  \Lambda\ \partial\bar\partial =0=
[\bar\delta +\frac12 \bar\partial \Lambda] \ \Lambda\ \partial\bar\partial\, .
\ee
These Bianchi identities seem to bear little relation to the gauge invariances~\eqn{gauge}. In fact, since we claim that our equations
are the natural generalization of Maxwell's equations $d^{*} d A=0$, the apparent asymmetry of our equations under adjoints
seems disappointing. Moreover, if $A$ is a general form, we also expect gauge for gauge symmetries analogous to those
for higher-form electromagnetism. In fact both these objections are easily remedied.

Firstly, if we multiply the K\"ahler equations\footnote{An analogous trick makes the symmetric tensor equations~\eqn{einst} self-adjoint.} on the left by the operator $$:I_{1}(2\sqrt{L\Lambda})/\sqrt{L\Lambda}:$$ where $:\bullet:$ denotes normal
ordering by form degree ({\it i.e.} with $L$ to the left and $\Lambda$ the right) and the modified Bessel function of the second kind obeys the asymptotic expansion\footnote{This series terminates in any given dimension when expanded in terms of $\Lambda$.}
$I_{1}(2\sqrt z)/\sqrt z=1+\frac12 z + \frac 1 {12} z^{2}+ \frac 1 {144} z^{3} + \frac1{2880} z^{4}+\cdots$, then we find the symmetric form 
for our equations
\be
G A=0
\ee
where the self-adjoint operator $G$ is given by 
\be
G=\  \scalebox{1.3}{:} \, \frac{I_{0}(2\sqrt{L\Lambda})}{\sqrt{L\Lambda}}\, (\Delta - 2\partial\delta-2\bar\partial\bar\delta)
+\ 2\, \frac{I_{1}(2\sqrt{L\Lambda})}{\sqrt{L\Lambda}} \, (\partial\bar\partial +\delta\bar\delta)
\, \scalebox{1.3}{:}\; ,
\ee
where all operators are normal ordered by form degree.
In turn, a gauge invariant action principle is $S=\int_{M}(A,GA)$.

We complete this Note by explicating the gauge for gauge symmetries of our system.
In the above symmetric formulation the gauge invariances and Bianchi identiies are
encapsulated by the operator identities
\be
\delta G = \bar\delta G = G  \partial=G\bar\partial=0\, .
\ee
If we introduce auxiliary, commuting variables $(p,\bar p)$ and the nilpotent operator
\be
D=\partial\frac{\partial}{\partial p} + \bar\partial\frac{\partial}{\partial \bar p}\; ,
\ee
then we can rewrite the gauge invariance~\eqn{gauge} as $A\sim A + D a$
where $a$ is any form linear in the auxiliary variables $(p,\bar p)$. However, since $D$
squares to zero on any (analytic) function of $(p,\bar p)$ we obtain a complex
$$
\cdots \stackrel{D}{\longrightarrow} {\cal F} \stackrel{D}{\longrightarrow} {\cal F}
\stackrel{D}{\longrightarrow} {\cal F}
\stackrel{D}{\longrightarrow}\cdots\, ,
$$
where ${\cal F}$ denotes forms depending on $(p,\bar p)$. This complex precisely encodes
the gauge for gauge symmetries of our K\"ahler equations.

The higher form K\"ahler electrodynamics announced in this Note were originally constructed by
BRST quantization of a K\"ahler spinning particle model (see~\cite{Marcus} for early studies). Similar K\"ahler systems and related higher form
equations of motion involving half of the SUSY generators have been constructed in~\cite{Fiorenzo}. A detailed description of these techniques will appear in a forthcoming paper~\cite{CLW}.
We thank Fiorenzo Bastianelli, Stefano Belluci, Roberto Bonezzi, Ori Ganor, Andy Neitzke, Boris Pioline, Martin Ro$\rm \check{c}$ek, Stanley Deser  and Bruno Zumino for stimulating discussions.

\end{document}